\documentclass[showpacs,amsmath,amssymb,aps,showkeys,floatfix,prd,a4paper]{revtex4}

\usepackage[dvips]{graphicx}
\usepackage{dcolumn}
\usepackage{bm}
\usepackage{epsfig}
\usepackage{amsfonts}
\usepackage{amssymb,amscd}

\def\lsim{\raise0.3ex\hbox{$<$\kern-0.75em\raise-1.1ex\hbox{$\sim$}}}
\def\gsim{\raise0.3ex\hbox{$>$\kern-0.75em\raise-1.1ex\hbox{$\sim$}}}

\def\pom{{I\!\!P}}

\def\beq{\begin{equation}}
\def\eeq{\end{equation}}
\def\bea{\begin{eqnarray}}
\def\eea{\end{eqnarray}}
\def\bq{\begin{quote}}
\def\eq{\end{quote}}

\def\gappeq{\mathrel{\rlap {\raise.5ex\hbox{$>$}}
{\lower.5ex\hbox{$\sim$}}}}

\def\lappeq{\mathrel{\rlap{\raise.5ex\hbox{$<$}}
{\lower.5ex\hbox{$\sim$}}}}

\def\Toprel#1\over#2{\mathrel{\mathop{#2}\limits^{#1}}}

\def\pom{{I\!\!P}}

\begin{document}


\title{Radion production in exclusive processes at CERN LHC}

\author{V.~P. Gon\c{c}alves}
\email{barros@ufpel.edu.br}
\author{W.~K. Sauter}
\email{werner.sauter@ufpel.edu.br}
\affiliation{High and Medium Energy Group, \\
Instituto de F\'{\i}sica e Matem\'atica, Universidade Federal de Pelotas\\
Caixa Postal 354, CEP 96010-900, Pelotas, RS, Brazil}
\date{\today}

\begin{abstract}
In the Randall-Sundrum (RS) scenario the  compactification radius of the extra dimension is stabilized by the radion, which is a scalar field lighter than the graviton Kaluza-Klein states. It implies that the detection of the radion will be the first signature of the stabilized RS model. In this paper we study the exclusive production of the radion in electromagnetic and diffractive hadron - hadron collisions at the LHC. Our results demonstrate that the diffractive production of radion is dominant and should be feasible of study at CERN LHC.

\end{abstract}


\maketitle
\section{Introduction}

The search  for particles beyond the Standard Model is one of the key issues of the ATLAS and CMS  experiments at LHC. In particular, these experiments could test the theories with extra dimensions, which aim to solve the hierarchy problem by bringing the gravity scale closer to the eletroweak scale (For a review see, e.g., \cite{hewett}). In recent years, the scenario proposed by Randall and Sundrum (RS)  \cite{rs}, in which there are two (3+1)-dimensional branes separated in a 5th dimension, has attracted a great deal of attention. This model predicts a Kaluza-Klein tower of gravitons and a graviscalar, called radion, which stabilize the size of the extra dimension without fine tuning of parameters and is the lowest gravitational excitation in this scenario. The mass of radion is expected to be of order of ${\cal{O}}$(TeV), which implies that the detection of the radion will be the first signature of the RS model.

Several authors have discussed the search of the radion in {\it inclusive} processes at Tevatron and LHC \cite{datta,bae,cheung,nayak,azuelos,battaglia,das}. In this paper we extend these previous studies for {\it exclusive} processes, in which the hadrons colliding remain intact after the interaction, losing only a small fraction of their initial energy and escaping the central detectors \cite{forshaw_review}.  The signal would be a clear one with a radion tagged in the central region of the detector accompanied by regions of low hadronic activity, the so-called "rapidity gaps". In contrast to the inclusive production, which is characterized by large QCD activity and backgrounds which complicate the identification of a new physics signal, the exclusive production will be characterized by a clean topology associated to hadron - hadron interactions mediated by colorless exchanges. In what follows, we will calculate the radion production considering   photon - photon or Pomeron - Pomeron interactions for $pp$, $pPb$ and $PbPb$ collisions at LHC energies. In particular, we will extend for radion production the perturbative KMR model \cite{kmr_first} (For a recent review see \cite{kmr_review}), which has been extensively used to calculate the exclusive diffractive production in hadron colliders considering Pomeron - Pomeron interactions, with predictions which are in agreement with the rates observed at the Tevatron \cite{cdf}.

This paper is organized as follows. In the next section we present a
brief review of the formalism necessary for calculate the radion production in electromagnetic and diffractive interactions in hadron - hadron collisions. Moreover, we discuss the extension of the survival probability gap for nuclear collisions. In Section \ref{res}  we present our results for the radion production in $pp$, $pA$ and $AA$ collisions for LHC energies. Finally, in Section \ref{sum} we present a summary of our main conclusions.

\section{Radion Production}
\label{radion}

\begin{figure}[t]
\begin{tabular}{cc}
\includegraphics[scale=0.35] {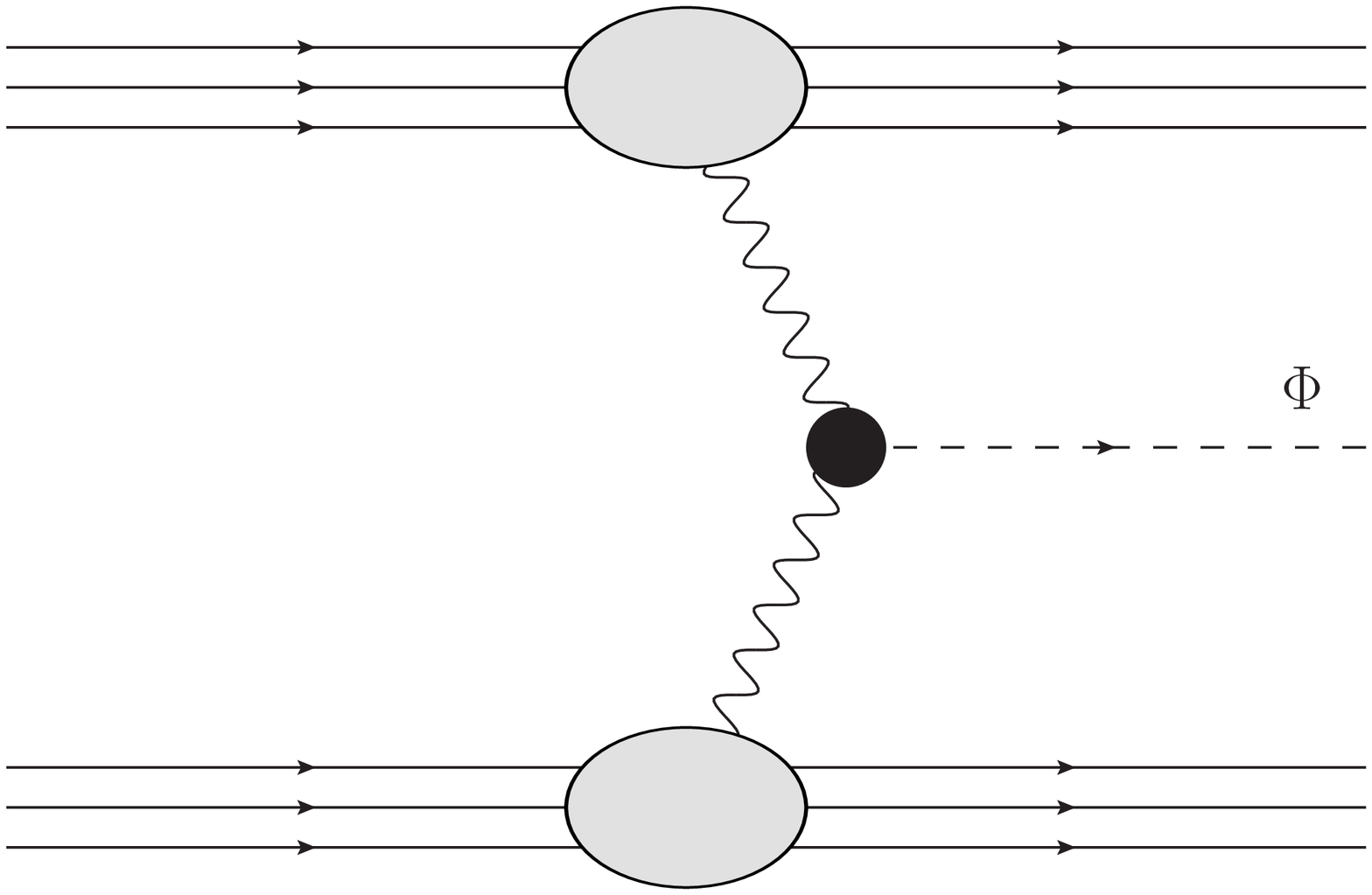} & \includegraphics[scale=0.35]{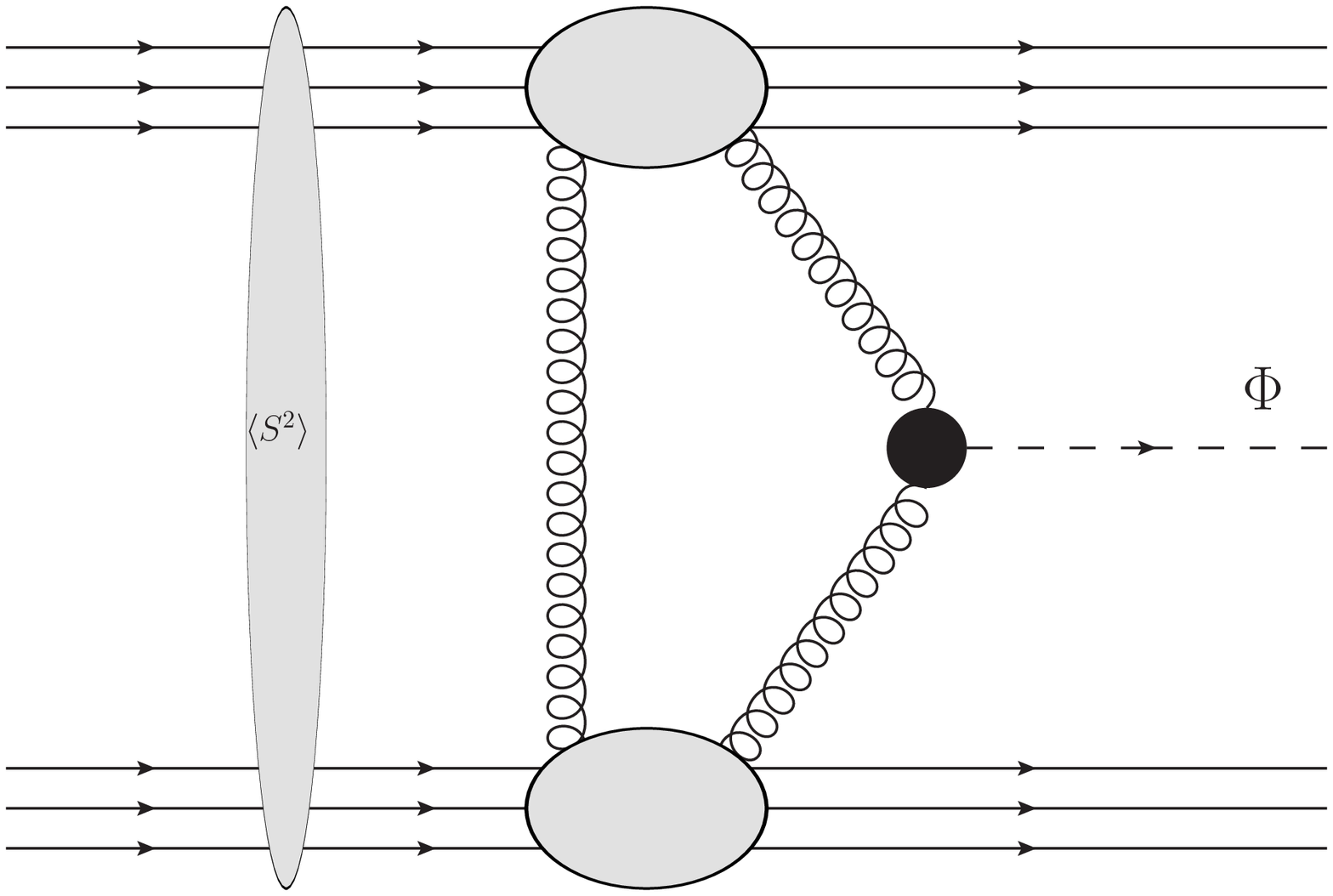} \\
(a) & (b)
\end{tabular}
\caption{(Color online) Exclusive radion production in (a) $\gamma \gamma$ and (b) Pomeron - Pomeron interactions.}
\label{fig1}
\end{figure}

In exclusive processes the radion can be produced in photon - photon and Pomeron - Pomeron interactions (See Fig. \ref{fig1}). In both cases, rapidity gaps between the radion and the hadrons colliding are expected.  {Let us} start our analysis considering the radion production in photon-photon interactions which can occur in coherent hadronic processes (For a review see, e.g., \cite{upcs}). At large impact parameter ($b > R_{h_1} +  R_{h_2}$) and ultra-relativistic energies, we expect  dominance of the  electromagnetic interaction. In  heavy ion colliders, the heavy nuclei give rise to strong electromagnetic fields due to the coherent action of all protons in the nucleus, which can interact with each other. Similarly, this also occurs when considering ultra- relativistic  protons in $pp(\bar{p})$ colliders. The photon emitted from the electromagnetic field
of one of the two colliding hadrons can interact with one photon of
the other hadron (two-photon process) or directly with the other hadron (photon-hadron
process). The total
cross section for a given process can be factorized in terms of the equivalent flux of photons of the hadron projectile and  the photon-photon or photon-target production cross section \cite{upcs}. In particular, the total cross section for the radion production will be given by
\begin{eqnarray}
 \sigma[h_1 h_2 \rightarrow h_1 \otimes \Phi \otimes h_2]  = \int_{0}^{\infty}\! \frac{d\omega_{1}}{\omega_{1}}\! \int_{0}^{\infty}\! \frac{d\omega_{2}}{\omega_{2}}\ F(\omega_1, \omega_{2})\, \sigma_{\gamma \gamma \rightarrow \Phi }(W_{\gamma\gamma} = \sqrt{4\omega_{1} \omega_{2}}), \label{eq:aa2axa}
\end{eqnarray}
where  $\otimes$ {means} the presence of a rapidity gap, $F(\omega_1, \omega_2)$ is the folded photon  spectra and $\sigma_{\gamma \gamma \rightarrow \Phi}$ is the cross-section of the fusion of two photons to produce the radion. The folded photon spectra is given by \cite{baur_ferreira}
\begin{eqnarray}
 F(\omega_1, \omega_{2}) = 2\pi \int_{R_{1}}^{\infty} db_{1} b_{1} \int_{R_{2}}^{\infty} db_{2} b_{2} \int_{0}^{2\pi} d\phi\ \, N_{1}(\omega_{1}, b_{1}) N_{2}(\omega_{2}, b_{2}) \Theta(b - R_{1} - R_{2}),
 \label{eq:fphotsp}
\end{eqnarray}
where $R_i$ are the projectile radii  and $b^2 = b_1^2 + b_2^2 - 2b_1 b_ 2 \cos \theta$. The theta function ensures that the hadrons do not overlap. The Weizs\"acker-Williams photon spectrum for a given impact parameter of two colliding hadrons is given by  \cite{upcs}
\begin{equation}
N(\omega,b) =\frac{\alpha_{em} Z^2}{\pi^2} \left(\frac{\xi}{b}\right)^2 \left\{ K_1^2(\xi) + \frac{1}{\gamma^2}  K^2_0(\xi)\right\} ,
\label{ene}
\end{equation}
with $K_{0,1}$ modified Bessel function of second kind, $\xi = \omega b/\gamma \beta$, $\beta$ is the speed of the hadron, $\gamma$ is the Lorentz factor and $\alpha_{em}$ is the electromagnetic coupling constant. 
This approximation is valid for heavy ion only, with the radii given by $R \simeq 1.2\,A^{1/3}$ fm. For  protons, the equivalent photon spectrum can be obtained from its elastic form factors in the dipole approximation (See e.g. \cite{david}). An alternative  is to use Eq. (\ref{ene}) assuming $R_p = 0.7$ fm for the proton radius, which implies a  good agreement with the parametrization of the luminosity obtained in \cite{Ohnemus:1993qw} for proton-proton collisions. The $\gamma \gamma \rightarrow \Phi$ cross section can be expressed as follows
\begin{equation}
 \sigma_{\gamma \gamma \rightarrow \Phi} = \frac{8 \pi^2}{m_{\phi}^3} \Gamma(\Phi \rightarrow \gamma\gamma) \,\,.
\end{equation}
The partial decay width of radion into two photons was calculated in \cite{bae,cheung} and is given by:
\begin{eqnarray}
 \Gamma(\Phi \rightarrow \gamma\gamma)   =   \frac{\alpha_{em}^2 m_{\Phi}^3}{256 \pi^3 \Lambda_{\Phi}^2} \left\lbrace  - \frac{22}{6} - [2 + 3x_W + 3x_W(2-x_W)f(x_W)] + \frac{8}{3} x_t [1+ (1-x_t)f(x_t)]  \right\rbrace ^2,
\end{eqnarray}
with $m_\Phi$ is the radion mass, $\Lambda_{\Phi} = \langle \Phi \rangle \approx {\cal{O}}(v)$ ($v$ the VEV of the Higgs field) determines the strength of the coupling of the radion to the standard model particles, $x_i = 4m_i^2/m_\Phi^2$ (with $i = W,t$ denoting the $W$ boson and the top quark) and the auxiliary function $f$ being given by
\begin{eqnarray}
f(z) = \left \{ \begin{array}{cr}
\left[ \sin^{-1} \left(\frac{1}{\sqrt{z}} \right ) \right ]^2\;, & z \ge 1 \\
-\frac{1}{4} \left[ \log \frac{1+\sqrt{1-z}}{1-\sqrt{1-z}} - i \pi \right ]^2
\;. & z <1 
\end{array}
\right . 
\end{eqnarray}

\begin{figure}[t]
\centerline{\psfig{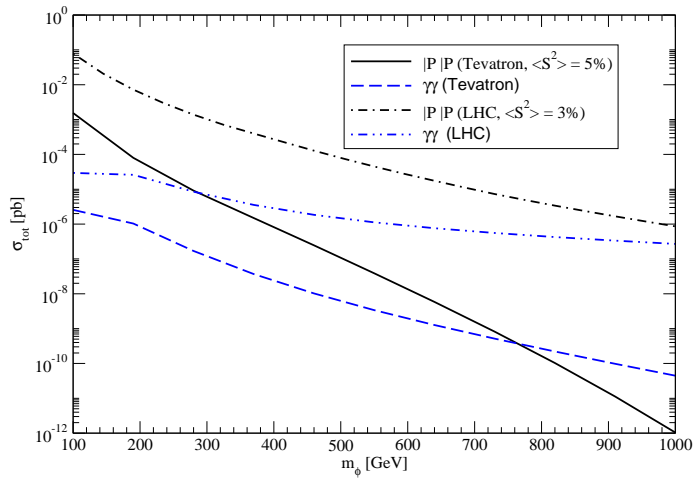}}
 \caption{(Color online) Total cross section for radion production in proton-proton collisions as function of radion mass for different mechanism of production, center of mass energies  and values of the survival probability $\langle {\mathcal S}^2 \rangle$.}
\label{fig2}
\end{figure}

One of the main channels for radion production in hadron colliders is the gluon fusion, $gg \rightarrow \Phi$, similarly to the SM Higgs boson production \cite{bae,cheung}. However, it gets further enhancement from the trace anomaly  for gauge fields, which implies that the  radion cross section is expected to be a factor $\approx$ 8 larger than the Higgs cross section over most of the mass range \cite{datta}. This result motivates the study of radion production in exclusive processes mediated by Pomeron - Pomeron ($\pom \pom$) interactions. 
A QCD mechanism for the exclusive diffractive production of a heavy central system has been proposed by Khoze, Martin and Ryskin (KMR) \cite{kmr_first} with its predictions in broad agreement with the observed rates measured by CDF Collaboration \cite{cdf} (For a recent study see Ref. \cite{forshaw_cor}). 
In this model, the total cross section for radion production can be expressed in a factorizated form as follows
\begin{equation}
 \sigma_{tot} = \int dy \langle {\mathcal S}^2 \rangle {\mathcal L}_{excl} \frac{2\pi^2}{m_{\Phi}^3} \Gamma(\Phi \rightarrow gg) 
\label{eq:kmr}
\end{equation}
where $\langle {\mathcal S}^2 \rangle$ is the survival probability gap (see below),  $\Gamma$ stand for the partial decay width of the radion $\Phi$ in a pair of gluons   and ${\mathcal L}_{excl}$ is the effective luminosity, given by
\begin{equation}
 {\mathcal L}_{excl} = \left[ {\cal{C}} \int \frac{dQ_t^2}{Q_t^4} f_g(x_1,x_1^{\prime},Q_t^2, \mu^2) f_g(x_2,x_2^{\prime},Q_t^2, \mu^2)\right]^2\,\,,
\end{equation}
where ${\cal{C}} = \pi/[(N_C^2 - 1)b]$, with $b$ the $t$-slope ($b = 4$ GeV$^{-2}$ in what follows), and the quantities $f_g$ being the  skewed unintegrated gluon densities.  Since $(x^{\prime} \approx Q_t/\sqrt{s}) \ll (x \approx M_{\Phi}/\sqrt{s}) \ll 1$ it is possible to express $f_g(x_1,x_1^{\prime},Q_t^2, \mu^2)$, to single log accuracy, in terms of the conventional integral gluon density $g(x)$, together with a known Sudakov suppression $T$ which ensures that the active gluons do not radiate in the evolution from $Q_t$ up to the hard scale $\mu \approx M_{\Phi}/2$.  Following  \cite{kmr_prosp} we will assume that
\begin{equation}
 f_g(x, Q_T^2, \mu^2) = R_g \frac{\partial}{\partial \ln Q_T^2} \left[ \sqrt{T(Q_t,\mu)}\ xg(x,Q_T^2) \right]
\end{equation}
where 
\[ R_g = \frac{2^{2\lambda+3}}{\sqrt{\pi}} \frac{\Gamma(\lambda + 5/2)}{\Gamma(\lambda + 4)} \]
accounts for the single $\log Q^2$ skewed effect, being $R_g \sim 1.2 (1.4)$ for LHC (Tevatron), and the Sudakov factor is given by
\begin{eqnarray}
T(Q_t,\mu)  =  \exp \left\{ -\int_{Q_t^2}^{\mu^2} \frac{dk_t^2}{k_t^2} \frac{\alpha_s(k_t^2)}{2\pi} \int_{0}^{1-\Delta} dz \, \left[ zP_{gg}(z) + \sum_{q} P_{qg} \right] \right\},
\end{eqnarray} 
with $\Delta = k_t/(\mu + k_t)$ and $P(z)$ the leading order DGLAP splitting functions. 
Moreover, the partial decay width of one radion into two gluons is \cite{bae,cheung}
\begin{equation}
 \Gamma[\Phi \rightarrow gg] = \frac{\alpha_s^2 m_{\Phi}^3}{32\pi^3 \Lambda_{\phi}^2} \left| \frac{15}{3} + x_t [1 + (1-x_t) f(x_t))] \right|^2 \label{radion:decay} \,\,.
\end{equation}
In this paper we will calculate  $f_g$ in the proton case considering that  the integrated gluon distribution $xg(x,Q_T^2)$ is described by the CTEQ6L parameterization \cite{cteq6}. In the nuclear case we will include the shadowing effects in $f_g^A$ considering that the nuclear gluon distribution is given by the EKS98 parameterization \cite{eks}, where  
$ x g_A(x,Q_T^2) = A R_g^A(x, Q_T^2) x g_p(x, Q_T^2)$, with $R_g$ describing the nuclear  effects in $xg_A$. Moreover, we assume in our calculations $\Lambda_{\phi} = 1$ TeV.



In order to obtain realistic predictions for the radion production by pomeron-pomeron fusion using the KMR model, it is crucial to use an adequate value for the survival probability gap, $\langle {\mathcal S}^2 \rangle$. This factor is the probability that secondaries, which are produced by soft rescatterings do not populate the rapidity gaps (For a detailed discussion see \cite{kmr_review}). In the case of proton-proton collisions, we will assume that $\langle {\mathcal S}^2 \rangle = 3 \, (5) \%$
for LHC (Tevatron) energies \cite{kmr_prosp}. However, the value of the survival probability for nuclear collisions still is an open question. An estimate of  $\langle {\mathcal S}^2 \rangle$ for nuclear collisions was calculated in  \cite{miller} using the Glauber model, which have obtained $\langle {\mathcal S}^2 \rangle = 8 \times 10^{-4} \,\,(8.16 \times 10^{-7})$ for pAu (AuAu) collisions at LHC energies. Another conservative estimate can be obtained assuming that  $\langle {\mathcal S}^2 \rangle_{A_1A_2} = \langle {\mathcal S}^2 \rangle_{pp}/(A_1.A_2)$ (For a discussion about nuclear diffraction see \cite{dif_nuc}). In what follows we will consider these two models for $\langle {\mathcal S}^2 \rangle_{A_1A_2}$ when considering radion production by pomeron-pomeron fusion. It is important to emphasize that, in contrast, the value of the  survival probability is of the order of unity for the $\gamma \gamma \rightarrow \Phi$ process in $pp/pA/AA$ collisions.

\section{results}
\label{res}

\begin{figure}[t]
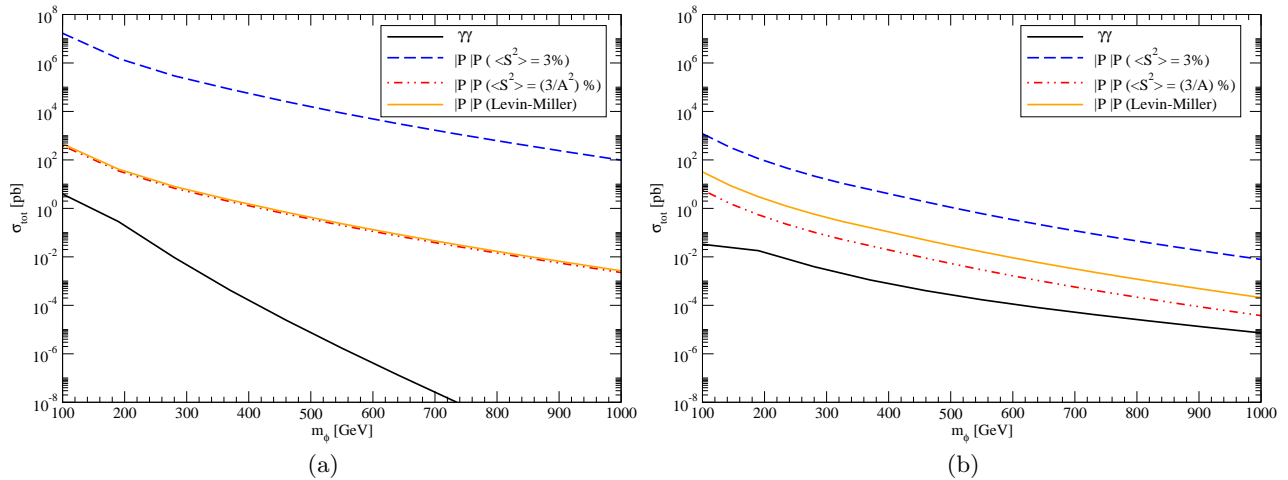

\begin{tabular}{cc}
\includegraphics[scale=0.35] {PbPb_mphi.eps} & \includegraphics[scale=0.35]{pPb_mphi.eps} \\
(a) & (b)
\end{tabular}
\caption{(Color online) Total cross section for radion production in (a) PbPb and (b) pPb collisions as function of radion mass for different mechanism of production and values of the survival probability $\langle {\mathcal S}^2 \rangle$.}
\label{fig3}
\end{figure}

In Figs. \ref{fig2} and \ref{fig3} we present our results for the dependence of the total cross section for radion production on the radion mass considering $\gamma \gamma$ and $\pom \pom$ interactions. In Fig. \ref{fig2} we calculate the cross section for $pp$ collisions and Tevatron ($\sqrt{s} = 2.0$ TeV) and LHC ($\sqrt{s} = 14.0$ TeV) energies. 
On the other hand, in Fig. \ref{fig3} we consider (a) $Pb Pb$ and (b) $pPb$ collisions for LHC energies: $\sqrt{s} = 5.5$ TeV and 8.8 TeV, respectively.   The $\gamma \gamma \rightarrow \Phi$ mechanism provides a natural lower limit for the radion exclusive production rate. Moreover, as  the photon flux scales as the squared charged of the beam, $Z^2$, two photon cross sections are extremely enhanced for ion beams. However, they are ever smaller than the predictions for radion production by  $\pom \pom$ interactions in $Pb Pb$ and $p Pb$ collisions, independently of the model considered for the survival probability $\langle {\mathcal S}^2 \rangle$. In the case of $pp$ collisions, they only are similar for large values of the radion mass. In comparison to the $pp$ case, the $\pom \pom$ predictions for $AA$ ($pPb$) nuclear collisions are enhanced by a factor $A^4$ ($A^2$). However, they are strongly reduced by the  survival probability. In the $Pb Pb$ case the predictions obtained using the model proposed in \cite{miller} and our naive model are very similar. In contrast, for $pPb$ collisions, they differ by a factor 5. Assuming the $\pom \pom$ predictions obtained using our model for $\langle {\mathcal S}^2 \rangle$ as being a lower bound and that $m_{\Phi} = 200$ GeV, one obtain that the radion production by this mechanism is a factor 180 (36) larger than the $\gamma \gamma$ predictions for $PbPb$   ($pPb$) collisions. In comparison to the exclusive Higgs production in $pp$ collisions \cite{kmr_prosp} we predict cross sections that are a factor $\approx 10$ larger. This enhancement is directly associated to the trace anomaly  for gauge fields, which leads to additional effective radion coupling to gluons or photons.

Let us  now  to compute the production rates for LHC energies  considering the distinct mechanisms.  The results are presented in Table \ref{tabradion}. At LHC we assume the  design luminosities ${\cal L} = 10^7 /\, 150 /\, 0.5$ mb$^{-1}$s$^{-1}$ for $pp/pPb/PbPb$  collisions at $\sqrt{s} = 14/\,8.8/\,5.5$ TeV and a run time of $10^7 \, (10^6)$ s for collisions with protons (ions). 
Moreover, we also consider the upgraded $pPb$ scenario proposed in Ref. \cite{david}, which analyse a potential path to improve the  $pPb$ luminosity and the running time. These authors proposed the following scenario for $pPb$ collisions:   ${\cal L} = 10^4$ mb$^{-1}$s$^{-1}$  and a run time of $10^7$ s. The corresponding event rates are presented in the third line of the  Table \ref{tabradion}.
Our results indicate that for the default settings and running times, the statistics are marginal for $PbPb$ collisions.  Consequently, the possibility to carry out a measurement of the radion in $\gamma \gamma$ and $\pom \pom$ interactions is virtually null in these collisions. On the other hand, in $pp$ collisions the event rates are reasonable, in particular for $\pom \pom$ interactions. In comparison to the inclusive radion production  \cite{cheung}, our predictions  for exclusive production are a factor $\le 10^{-4}$  smaller.  Despite their much smaller cross sections, the clean topology of exclusive radion production implies a larger signal to background ratio. Therefore, the experimental detection is in principle feasible. However, the signal is expected to be reduced due to the event pileup which eliminates one of the main advantages  of the exclusive processes.
In contrast, in $pA$ collisions it is expected to trigger on and carry out the measurement with almost no pileup \cite{david}. Therefore, the upgraded $pA$ scenario provides the best possibility to detect the radion in an exclusive process.

\begin{table}
\begin{center}
\begin{tabular} {||c|c|c||}
\hline
\hline
& $\gamma \gamma$ &  $\pom \pom$  \\
\hline
\hline
$pp$ &  $3.0$  & $700$  \\
\hline
$pPb$ & $4.5 \times 10^{-3}$ & $9 \times 10^{-2}$  \\
 & 3 & 60 \\
\hline
$PbPb$& $2.5 \times 10^{-4}$ & $3.5 \times 10^{-1}$ \\
\hline
\hline
\end{tabular}
\end{center}
\caption{ The events rate/year for the  radion production  in $pp/pPb/PbPb$  collisions at LHC energies considering the $\gamma \gamma$ and $\pom \pom$ mechanisms and $m_{\Phi} = 200$ GeV.}
\label{tabradion}
\end{table}

Some  comments are in order. The exclusive cross sections for radion production are inversely proportional to $\Lambda_{\Phi}^2$, which still is a free parameter.  Here we assume $\Lambda_{\Phi} = 1$ TeV as in  \cite{bae,cheung}. Moreover, our predictions are strongly dependent of the radion mass. In Table \ref{tabradion} we have considered  $m_{\Phi} = 200$ GeV. Larger values imply that the radion measurement in exclusive processes  would not be possible at LHC. 
Finally, the value of the survival probability for  processes involving nuclei still is an open question. In this paper we  calculated the cross sections considering two phenomenological models for $\langle {\mathcal S}^2 \rangle$ and estimated the event rates assuming a pessimistic scenario. However, these points deserve a more detailed study which we postpone for a future publication.

\section{Summary}
\label{sum}

Exclusive processes have already been observed at the Tevatron with rates in broad agreement with the theoretical predictions.  It is expected that at LHC the addition of forward proton taggers \cite{f420} to enhance the discovery and physics potential of the ATLAS, CMS and ALICE detectors \cite{cms,atlas,alice}. One of the possible scenarios which could be analyzed is that  proposed by Randall and Sundrum (RS)  \cite{rs},  which predicts the radion as  the lowest gravitational excitation in order to stabilize the size of the extra dimension. As the mass of radion is expected to be $\approx$ TeV,  the detection of the radion will be the first signature of the RS model. In this paper we study the exclusive production of the radion in electromagnetic and diffractive hadron - hadron collisions at the LHC. We predict larger cross sections in comparison to the Higgs production in exclusive processes. Our results demonstrate that the diffractive production of radion is dominant and should be feasible of study at CERN LHC.

\section*{Acknowledgements}
 This work was partially financed by the Brazilian funding agencies CNPq and CAPES.



\end{document}